\documentclass{PoS}
\title{Worldline Approach to Chiral Fermions }

\ShortTitle{Chiral Fermions}

\author{\speaker{Kurt Langfeld} \\  % \thanks{A footnote may follow.}\\
        School of Maths \& Stats, University of Plymouth, \\
        Plymouth, PL4 8AA, England \\
        E-mail: \email{Kurt.Langeld@plymouth.ac.uk}}

\author{Gerald Dunne \\
        Department of Physics,  University of Connecticut, \\
        Storrs, CT 06269-3046, USA \\
        E-mail: \email{dunne@phys.uconn.edu}}

\author{Holger Gies \\
        Institut f\"ur Theoretische Physik, Universit\"at  Heidelberg \\
        D-69120 Heidelberg, Germany  \\
        E-mail: \email{h.gies@thphys.uni-heidelberg.de}}

\author{Klaus Klingm\"uller \\
        Institut f\"ur Theoretische Physik, Universit\"at  Heidelberg \\
        D-69120 Heidelberg, Germany  \\
        E-mail: \email{k.klingmueller@thphys.uni-heidelberg.de}}

\abstract{We propose to apply ``worldline numerics'' to a numerical
calculation of quark determinants. The Gross-Neveu model with a U(1)
chiral symmetry is considered as a first test.  The worldline approach
allows for an analytic renormalisation, and only finite parts of the
determinant require a numerical calculation. It is shown that the
discretisation of the worldlines, which is central to the numerical
treatment, preserves chiral symmetry exactly. Numerical results for a
kink configuration as a scalar background field are shown and compared
with analytical results.  The case of finite fermion chemical
potential is also briefly discussed.  }

\FullConference{The XXV International Symposium on Lattice Field Theory\\
		 July 30-4 August 2007\\
		 Regensburg, Germany}

\newcommand{\be}{\begin{equation}}
\newcommand{\en}{\end{equation}}
\newcommand{\bea}{\begin{eqnarray}}
\newcommand{\ena}{\end{eqnarray}}

\newcommand{\hbo}{\hbox to 1 true cm {\hfill } }
\newcommand{\tr}{\hbox{tr}}

\def\dslash{\partial\kern-.6em\slash}

\begin{document}

\section{Introduction:}

Over the last decade, the QCD phase diagram as a function of temperature
and baryon density has attracted intense investigations using
computer simulations and collider experiments such as
RHIC undertaken at the Brookhaven National Laboratory.
From the simulation point of view, many efforts in lattice gauge theory
in the recent past were devoted to control the severe sign problem.
The proposals use a Taylor expansion with respect to the
baryon chemical potential~\cite{Choe:2002mt,Allton:2002zi,Ejiri:2003dc},
imaginary values for the chemical potential
$\mu $~\cite{Alford:1998sd,de Forcrand:2002ci,D'Elia:2002gd} or
overlap enhancing techniques~\cite{Fodor:2001au%,Fodor:2001pe%
%,Fodor:2004nz
}.
Despite of these successes, our knowledge of the QCD phase diagram
is still limited to rather small values of the chemical potential.

Further insights have come from QCD-like theories
such as 2-colour QCD~\cite{Hands:1999md,Kogut:2002cm,Hands:2006ve},
from perturbative studies and QCD motivated models.
In the latter case, one assumes that the phase at highest densities
is homogeneous and argues that quark matter is organised in a colour
superconducting
state~\cite{Bailin:1983bm,Alford:1997zt,Schafer:1999jg,Pisarski:1999tv}.
Recent studies of the Gross Neveu model in the limit of many flavours
have attracted a lot of interest since it was found that
the high-density state of fermion matter forms an inhomogeneous
``baryon crystal''~\cite{Thies:2003kk%,Thies:2006ti,Boehmer:2007ea
}.

Good chiral properties of the fermion action is of central importance
for an investigation of quark matter at intermediate densities, since
the high-density transition is driven by chiral dynamics.
Unfortunately, lattice fermion actions necessarily suffer from the
fermion doubling problem as firstly pointed out by
Nielsen and Ninomiya~\cite{Nielsen:1980rz}. Nowadays, {\it staggered
fermions}~\cite{Kogut:1974ag}, {\it domain wall fermions}~\cite{Kaplan:1992bt}
or {\it Neuberger fermions}~\cite{Neuberger:1997fp}, which are an explicit
realisation of the Ginsparg-Wilson relation~\cite{Ginsparg:1981bj},
are widely used in numerical simulations. Despite of these advanced
formulations and great numerical efforts, it turns out cumbersome to
achieve good chiral properties such as a sufficiently small pion mass.

Since the {\em worldline approach} to the quark determinant does not use
a lattice discretisation of space-time, it circumvents many of these
significant difficulties. Here, we will argue that the prospects
of the worldline approach are (i) exact chiral symmetry but yet
a fully numerical approach, (ii) analytic renormalisation and (iii)
a clear description of Fermi surface effects.
%Note that up to now only indirect evidence of Fermi surface effects
%were obtained in numerical simulations of effective quark
%models~\cite{Hands:2001ck,Hands:2003dh}.

The worldline method is a string-inspired approach to quantum
field theory; see~\cite{Schubert:2001he} for a review.
It was further developed into a viable tool for an efficient calculation
of functional determinants for arbitrary background
fields~\cite{Gies:2001zp%,Gies:2001tj
}. Subsequently,
{\em worldline numerics} has enjoyed a wide span of
applications ranging from the Casimir
effect~\cite{Gies:2003cv,Gies:2006bt%,Gies:2006cq,Gies:2006xe
} and
fermion induced quantum interactions~\cite{Langfeld:2002vy} to
the description of pair production in inhomogeneous
fields~\cite{Dunne:2005sx%,Dunne:2006st,Gies:2005bz
}. A worldline lattice formulation has been presented in
\cite{Schmidt:2003bf}.

\section{The chiral Gross-Neveu model }

\subsection{Setup of the model }

The Gross-Neveu model in its original formulation is a two dimensional
fermionic theory which shares with QCD the property of spontaneous
chiral symmetry breaking and asymptotic freedom~\cite{Gross:1974jv}.
Due to Thies and Urlichs, the phase diagram is analytically known in
the limit of many flavours $N$ \cite{Thies:2003kk%,Thies:2006ti
}.  This
model therefore provides for a benchmark test for any new numerical method
which tries to extend its reach to very dense fermionic systems.

%\vskip 0.3cm
In the chiral version of this model, a pseudo-scalar
field $\pi (x)$ acts as chiral partner of the scalar field
$\sigma (x)$. The partition function is given by
\bea
{\cal Z } &=& \int {\cal D}  \sigma  \, {\cal D}
\pi \, \exp
\Bigl\{ - \, N \, S_\mathrm{fer} \, - \,  N \, S_\mathrm{bos} \Bigr\} \; ,
\hbo
 S_\mathrm{bos} = \frac{1}{2g^2} \int d^2x \,
\Bigl[ \sigma  ^2(x) \, + \, \pi ^2(x) \Bigr] \, ,
\label{eq:1} \\
S_\mathrm{fer}  &=& \frac{1}{2} \, \hbox{tr} \, \ln \,
\Bigl(
- \, \partial ^2 \, + \, \sigma ^2 + \, \pi ^2
\, - \, i \,
\dslash \sigma  \, + \, \gamma _5 \,
\dslash \pi \Bigr) \; ,
\label{eq:2}
\ena
where $g$ is the bare coupling constant and
where we have used anti-hermitian Dirac matrices.
The partition function is invariant under a $U(1)$ chiral rotation of
the fields:
\be
\left( \begin{array}{c} \sigma ^\prime (x) \\ \pi ^\prime(x)
\end{array} \right) \; = \;
\left( \begin{array}{cc} \cos \theta & - \sin \theta \\
\sin \theta & \cos \theta \end{array} \right) \;
\left( \begin{array}{c} \sigma (x) \\ \pi (x) \end{array} \right) \; .
\label{eq:3}
\en
In the large-$N$ limit, one assumes that fluctuations of the mesons
are negligible, and that the relevant field configurations can be obtained
in leading-order saddle point approximation:
\be
S_\mathrm{fer} \, + \, \int d^2x \,
\frac{N}{2g^2} \Bigl[ \sigma  ^2(x) \, + \, \pi ^2(x) \Bigr]
\; \stackrel{ \sigma, \pi }{\longrightarrow } \; \mathrm{min}.
\label{eq:N}
\en

\subsection{ The worldline approach to the GN model }

The key ingredient of the worldline calculation of the fermionic part
$S_\mathrm{fer}$ in (\ref{eq:2}) is the representation of
$S_\mathrm{fer}$ in terms of an ensemble average of closed loops
$x_\mu(\tau )$, $\tau =0 \ldots T $ , $x_\mu(0) = x_\mu(T)$, in Euclidean
spacetime.  In the {\em loop cloud}
approach~\cite{Gies:2001zp%,Gies:2001tj
}, the worldlines are generated
according to the free probabilistic measure: \be \delta \Bigl(
x_\mathrm{cm}[x] - x_{\mathrm{c}} \Bigr) \; \exp \left\{ - \; \int _0^T d\tau \;
  \left[ \frac{\dot{x}^2}{4} \right] \; \right\} \; ,
\label{eq:4}
\en
where the loop centre of mass given by
\be
x_\mathrm{cm}[x] \; = \; \frac{1}{T} \int _0^T d\tau \; x(\tau )
\label{eq:5}
\en
is constrained to $x_{\mathrm{c}}$. The fermion determinant is then represented
by
\bea
S_\mathrm{fer} &=& \frac{1}{8\pi } \int _{1/\Lambda ^2} ^\infty
\frac{dT}{T^2} \; \int d^2x_{\mathrm{c}} \; \left\langle
\exp \left\{ - \int _0^T d\tau \, \left( \sigma ^2 + \pi ^2 \right)
\right\} \; \Gamma (\sigma , \pi) \; \right\rangle \; ,
\label{eq:6} \\
&&\Gamma (\sigma , \pi) =
\tr_\gamma \; {\cal P} \, \hbox{exp} \left( i \, \int _0^T d\tau \;
 \left( \dslash \sigma \, + \, i  \, \gamma _5 \,
\dslash \pi \right) \, \right) \; ,
\label{eq:8}
\ena
where $\Lambda$ is a UV cutoff.

\subsection{Exact chiral symmetry }

In numerical calculations, a closed loop $x(t)$ is represented by
a finite number of points:
$$
x_i \rightarrow x(t_i), \hbo  i=1 \ldots N_{\mathrm{p}} , \hbo
d\tau = T/N_{\mathrm{p}} \; .
$$
The spin factor $\Gamma (\sigma , \pi)$ is approximated by
a path-ordered product
\be
\Gamma _\mathrm{dis}(\sigma , \pi) \; = \; \tr _\gamma \; \prod _{x_i}
{\cal P} \,
\hbox{exp} \left( i \, d\tau \;
 \left[ \dslash \sigma (x_i) \, + \, i  \, \gamma _5 \,
\dslash \pi (x_i) \right] \, \right) \; ,
\label{eq:9}
\en
The crucial observation is that, in spite of the discretisation,
$\Gamma _\mathrm{dis}$ still is exactly chirally invariant.
To show this, we define a unitary matrix $U$ by
$
U \; = \; \cos (\theta/2) \, + \, i \sin (\theta/2) \; \gamma _5 \; ,
$
and show that
\be
\dslash \sigma ^\prime (x) \, + \, i  \, \gamma _5 \,
\dslash \pi^\prime (x)  \; = \;
U \; \left(
\dslash \sigma  (x) \, + \, i  \, \gamma _5 \,
\dslash \pi  (x) \right) \; U^\dagger.
\label{eq:10}
\en
Because of the path ordering and the closeness of the (discretised) loops,
we easily find that (see figure~\ref{fig:1}, left panel for an
illustration)
$$
\Gamma _\mathrm{dis}(\sigma ^\prime, \pi^\prime) \; = \;
\Gamma _\mathrm{dis}(\sigma , \pi) \; .
$$
The other parts of the fermionic action (\ref{eq:6}) as well as the
integration measure for the mesonic fields are trivially invariant
(there is no anomaly in this model)
leaving us with an exact chiral symmetry for the discretised theory.

\subsection{Renormalisation }

Another big advantage of the worldline approach to fermionic
determinants is that the UV regularisation can be performed along the
lines made explicit in the ab initio continuum formulation.
Only finite parts of the determinant must be calculated by numerical means.
This implies that one does not need to invoke any ``order-$a$''
improvement which is instrumental when conventional lattice fermions
are considered.
Let us illustrate the renormalisation procedure for the present case.
Introducing the space-time average
\be
M^2 \; = \; \frac{1}{L^2} \int d^2x \; [ \sigma ^2 (x) + \pi ^2 (x) ],
\label{eq:50}
\en
the fermionic action can be split into a UV divergent and a finite
part:
\bea
S_\mathrm{fer} &=& S_\mathrm{0}(M, \Lambda) \; + \;
S^\mathrm{fin}_\mathrm{fer} [\sigma, \pi ]\; ,
\label{eq:51} \\
 S^\mathrm{fin}_\mathrm{fer} [\sigma, \pi ] &=&
\frac{1}{8\pi } \int _{0} ^\infty
\frac{dT}{T^2} \; \int d^2x \; \Bigl\langle
\exp \left\{ - \int _0^T d\tau \, (\sigma ^2 + \pi ^2 )\right\} \;
\tr _\gamma \; {\cal P} \, \hbox{exp} \left( i \, \int _0^T d\tau \;
( \dslash \sigma \, + \, i \gamma _5 \, \dslash \pi ) \right)
\nonumber \\
&&- 2 \exp \left\{ - T \,M^2 \right\} \;
\Bigr\rangle _x \; ,
\label{eq:52} \\
S_\mathrm{0} (M, \Lambda) &=& \frac{L^2}{4\pi } \int _{1/\Lambda ^2} ^\infty
\frac{dT}{T^2} \; \exp \left\{ - T \, M^2 \right\} \; .
\label{eq:53}
\ena
With this construction, the part $ S^\mathrm{fin}_\mathrm{fer}$ of the action
which involves time consuming numerical simulations is UV and IR finite.
Accordingly, we have removed the regulator in (\ref{eq:52}) by taking the limit
$\Lambda \to \infty $.
The part $S_\mathrm{0}$ of the action contains the divergent pieces
which can be calculated explicitly; dropping a field-independent
constant, we obtain
\be
S_\mathrm{0} (M, \Lambda) \; = \;  \frac{L^2}{4\pi } \Bigl[ \;
M^2 \; \ln \frac{M^2}{\Lambda^2} \; + \; (\gamma_E -1) \; M^2 \Bigr]
\; + \; {\cal O} \left( \frac{M^2}{\Lambda ^2} \right) \; ,
\label{eq:54}
\en
where $\gamma _E $ is Euler's constant. Adding the bare bosonic part of the action
in (\ref{eq:1}), we can impose renormalization conditions, for
instance, of Coleman-Weinberg type; this defines the renormalized coupling
at an RG scale $\mu$, $g^{-2}(\mu):= \partial^2 S/\partial
\sigma^2|_{M^2=\sigma^2=\mu^2}$, finally yielding,
\be
S_\mathrm{0} (M, \Lambda) \; + \; S_\mathrm{bos} \; = \;
\frac{L^2}{4\pi } \; M^2 \; \left( \ln \frac{M^2}{M_0^2} -1 \right) \; ,
\hbo
M_0^2= \mu^2 \; e^2\; e^{-2\pi/g^2(\mu)},
\en
where we have traded the coupling $g(\mu)$ for an RG invariant
mass scale $M_0$ in the large $N$ limit, reflecting
dimensional transmutation. This scale also denotes the large-$N$
minimum of the action at zero temperature and density,
$\sigma^2=M_0^2=$const.

\subsection{A numerical benchmark test }

\begin{figure}[t]
\includegraphics[height=5cm]{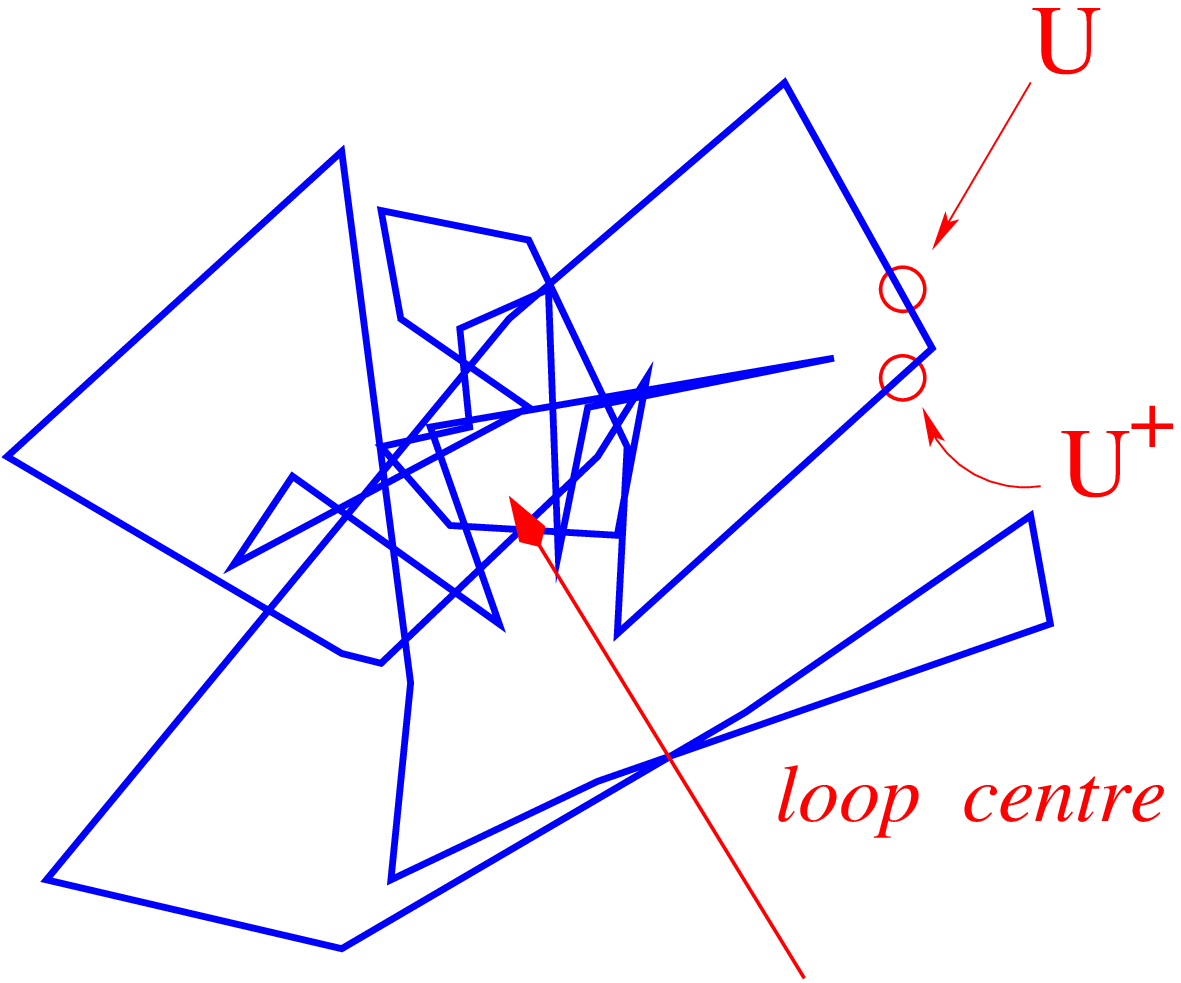}
\includegraphics[height=7.5cm]{ShapeOfKinkZeroMode.eps}
\caption{ Illustration of chiral invariance of the spin factor (left);
zero-mode wave function reconstructed from the heat
kernel  using free loop ensembles (right). \label{fig:1} }
\end{figure}
For a benchmark test, we choose a kink configuration as a
background field:
$$
\sigma (t,x) \; = \; \sigma (x) = \hbox{tanh}(x) \; , \hbo
\pi (t,x) \; = \; 0 \; .
$$
This kink interpolates between the two homogeneous vacuum states $
\sigma = \pm 1$ and is the basic building block of the \lq baryonic
crystal\rq \, of the Gross-Neveu model~\cite{Thies:2003kk}.  The Dirac
structure decomposes into two Schr\"odinger problems for the
heat-kernel traces
\be
\tr_x \exp \{ - T  H_\pm \} = \frac{1}{\sqrt{4\pi T}} \Bigl\langle \exp
\{ - T  H_\pm \} \Bigr\rangle _{ x } \; , \hbo
H_\pm \; = \; - \partial ^2 + \sigma ^2 \pm \frac{d}{dx} \sigma (x).
\label{eq:11}
\en
On the other hand, the heat kernel expectation value can be
expressed in terms of the eigenmodes of the Hamiltonian $H_\pm$. In
particular, $H_-=   - \partial ^2 + 1 - 2/\mathrm{cosh} ^2 x$ gives
rise to a zero mode,
\be
\frac{1}{\sqrt{4\pi T}}\Bigl\langle \exp \{ - T  H_- \} \Bigr\rangle _{ x }
\, = \, \vert \psi _0  (x) \vert^2 + \sum _i
\vert \psi _i  (x)\vert ^2
\, \mathrm{e}^{ - T E_i } \, .
\label{eq:12}
\en This choice therefore challenges the worldline approach, and the
crucial question is whether a moderate number of free loops is able to
grasp the zero-mode contribution.  We have calculated the heat kernel
expectation value for a range of propertime values $T$ of
$\mathcal{O}(10)$ using 50000 loops consisting of $100$ points per
loop, all of which include the point $x$.  In this propertime range,
the contributions of the excited states is small, and the modulus of
the zero-mode wavefunction $\vert \psi _0 \vert ^2 $ can be
reconstructed from (\ref{eq:12}) by a fit.  Our numerical findings for
the modulus of the zero mode wavefunction are compared in
figure~\ref{fig:1}, right panel, with the exact result.  A rather
rough discretisation of the loops already yields quite accurate
results.

\section{Finite densities }

Let us consider the case of a time independent, but non-homogeneous
scalar field ($\pi $ is set to zero for illustration purposes),
which is relevant for the high density crystal phase.
The fermion determinant at finite temperatures $1/\beta $ and
finite fermion chemical potential $\mu $ in the worldline
approach is given by
\bea
S_\mathrm{fer} &=& \frac{1}{2} \int _{1/\Lambda ^2} ^\infty
\frac{dT}{T} \; \sum_n \; \; \exp \left\{ - T \, \left[
\frac{2 \pi}{ \beta } \, (n+1/2) \, - \, i \mu \right]^2 \; \right\}
\; K(T) \; ,
\label{eq:20} \\
K(T) &=& \frac{1}{\sqrt{4\pi \, T }} \; \int dx_{\mathrm{c}} \; \left\langle
\exp \left\{ - \int _0^T d\tau \, \sigma ^2 \right\} \;
\tr _\gamma \; {\cal P} \, \hbox{cosh}
\left( i \, \int _0^T d\tau \; \dslash \sigma \, \right)
\; \right\rangle \; .
\label{eq:21}
\ena
In order to make the physics of the Fermi surface transparent,
we introduce the Laplace transform of the kernel $K(T)$ by
\be
K(T) \; = \; 2 \int_0^\infty dE \; E \; \exp [ - T E^2 ] \; \rho (E) \; ,
\label{eq:22}
\en
where $\rho (E)$ has the interpretation of the density of states.
We obtain:
\be
S_\mathrm{fer} \; = \; 2 \int dE \; E \; \rho (E) \;
\frac{1}{2} \int _{1/\Lambda ^2} ^\infty
\frac{dT}{T} \; \sum_n \; \; \exp \left\{ - T \, \left[
\frac{2 \pi}{ \beta } \, (n+1/2) \, - \, i \mu \right]^2 \;
- \; T \, E^2 \right\} \; .
\label{eq:23}
\en
The technical advantage is that we have mapped the problem of dealing
numerically with the Fermi surface onto the problem of a free particle
theory with single particle energy $E$. Hence, it is well known how to
evaluate the proper time integration $T$ and the Matsubara sum $n$ in
(\ref{eq:23}). Decomposing the fermionic action into temperature
dependent and independent parts, we find:
\bea
S_\mathrm{fer} &=& S_\mathrm{fer}^{\mathrm{temp}}
\; + \; S_\mathrm{fer}^{0}
\label{eq:24} \\
S_\mathrm{fer}^{\mathrm{temp}} &=& \int dE \; E \; \rho (E) \; \biggl\{
\ln \Bigl[ 1 \; + \; \exp \{ - \beta (E+\mu) \} \Bigr] \; + \;
\ln \Bigl[ 1 \; + \; \exp \{ - \beta (E-\mu) \} \Bigr] \; \biggr\}  .
\label{eq:25} \\
S_\mathrm{fer}^{0}  &=& \int dE \; E \; \rho (E) \;
 \frac{1}{2} \int _{1/\Lambda ^2} ^\infty
\frac{dT}{T} \; \beta \; \frac{dk_0}{2 \pi} \;
\; \exp \left\{ - T \, \left[ k_0^2 \, + \, E^2 \right]^2
\right\} \; .
\label{eq:26}
\ena
Hence, the formulation offers a complete control over the physics associated
with the Fermi surface, and even the low temperature and high density
regime is accessible. For instance, the small temperature expansion
(arbitrary chemical potential) of the baryon density is given by
\be
\frac{1}{\beta } \frac{d}{d\mu } \; S_\mathrm{fer}^{\mathrm{temp}} \; = \;
\int _0^\mu dE \; E \, \rho (E) \; + \; \frac{\pi^2}{6} T^2 \;
\frac{d}{dE} \, [E \rho (E)] \, \vert _{E=\mu} \; + \; {\cal O}(T^4) \; .
\label{eq:27}
\en
Fermi surface effects can thus be studied in a systematic fashion.
Note, however, that the numerical calculation of the
density of states $\rho(E)$ from the kernel $K(T)$ in (\ref{eq:22})
can be cumbersome.

\end{document}